\documentclass[08pt,twocolumn,preprintnumbers,amsmath,amssymb,nofootinbib,superscriptaddress,pra]{revtex4-1}

\usepackage{amsmath,amssymb,graphicx}
\usepackage{float}
\usepackage{latexsym} 
\usepackage{amsmath,amsthm}
\usepackage{ifpdf}
\usepackage{epstopdf}
\usepackage{color}
\usepackage{epsfig}
\usepackage{dcolumn}
\usepackage{bm}
\usepackage{braket}
\usepackage[symbol]{footmisc}
\usepackage{soul}
\usepackage{xcolor}
\usepackage{graphicx}

\begin{document}
\def \beq{\begin{equation}}
\def \eeq{\end{equation}}
\def \bea{\begin{eqnarray}}
\def \eea{\end{eqnarray}}
\def \bes{\begin{split}}
\def \ees{\end{split}}
\def \besu{\begin{subequations}}
\def \esu{\end{subequations}}
\def \bea{\begin{align}}
\def \eal{\end{align}}
\def \bem{\begin{displaymath}}
\def \eem{\end{displaymath}}
\def \P{\Psi}
\def \Pd{|\Psi(\boldsymbol{r})|}
\def \Pds{|\Psi^{\ast}(\boldsymbol{r})|}
\def \Po{\overline{\Psi}}
\def \bs{\boldsymbol}
\def \dert{\frac{d}{dt}}
\def \k{\ket}
\def \br{\bra}
\def \bm{\hat b^-_{\Omega}}
\def \bp{\hat b^+_{\Omega}}
\def \am{\hat a^-_{\omega}}
\def \ap{\hat a^+_{\omega}}
\def \pau {\partial_u}
\def \pav{\partial_v}
\def \paut{\partial_{\tilde u}}
\def \pavt{\partial_{\tilde v}}
\def \a{\alpha_{\Omega \omega}}
\def \b{\beta_{\Omega \omega}}

\title{Penrose Superradiance in Nonlinear Optics}
\author{Maria Chiara Braidotti}
\affiliation{School of Physics and Astronomy, University of Glasgow, G12 8QQ, Glasgow, United Kingdom}
\author{Daniele Faccio}
\affiliation{School of Physics and Astronomy, University of Glasgow, G12 8QQ, Glasgow, United Kingdom}
\affiliation{Wyant College of Optical Sciences, University of Arizona, Tucson, Arizona 85721, USA}
\author{Ewan M. Wright}
\affiliation{Wyant College of Optical Sciences, University of Arizona, Tucson, Arizona 85721, USA}
\email{ewan@optics.arizona.edu; daniele.faccio@glasgow.ac.uk; MariaChiara.Braidotti@glasgow.ac.uk}

\begin{abstract}
\noindent Particles or waves scattered from a rotating black hole can be amplified through the process of Penrose superradiance, though this cannot currently be observed in an astrophysical setting. However, analogue gravity studies can create generic rotating geometries exhibiting an ergoregion, and this led to the first observation of Penrose superradiance as the over-reflection of water waves from a rotating fluid vortex. Here we theoretically demonstrate that Penrose superradiance arises naturally in the field of nonlinear optics. In particular, we elucidate the mechanism by which a signal beam can experience gain or amplification as it glances off a strong vortex pump beam in a nonlinear defocusing medium.  This involves the trapping of negative norm modes in the core of the pump vortex, as predicted by Penrose, which in turn provides a gain mechanism for the signal beam. Our results elucidate a new regime of nonlinear optics involving the notion of an ergoregion, and provide further insight into the processes involved in Penrose superradiance.
\end{abstract}

%\pacs{04.70.Dy, 04.62.+v, 04.70.-s, 97.60.Lf, 04.62.+v, 04.60.-m, 05.45.Yv}% PACS, the Physics and Astronomy
                             % Classification Scheme.
%\keywords{Black hole, Haw\-king radiation, sine-Gordon equation, quantum soliton evaporation, AKNS system, geometrization, soliton quantization.}%Use showkeys class option if keyword
                              %display desired

\maketitle

{\emph{Introduction.}}
Penrose or rotational superradiance is a process in which waves scattered from a rotating black hole can extract energy from it. In 1969 Penrose predicted this effect noticing that for an asymptotic observer, particles that fall inside  the ergoregion around a rotating Kerr black hole will have negative energy and thus lead to amplification of a reflected positive energy component \cite{Penrose1969}. %Interaction with a Kerr black hole can therefore lead to a gain in energy at the expense of the rotational energy of the black hole itself \cite{Penrose1969}. 
This concept was later
%investigated as a means of energy mining from a black hole \cite{mining1,mining2} and
extended by Zel'dovich to the prediction of amplification of waves reflected from a rotating, metallic (i.e. absorbing) cylinder \cite{Zeldy1971,goodingReinventing2019,MC}, with recent studies \cite{Silke_sound,Ewan19,goodingDynamics2020} and an experimental demonstration using sound waves \cite{Marion}.\\
%
%Measuring this astrophysical process is currently well  beyond current technology due to the enormous distances at play. 
In the last decades, analogue gravity studies have attracted considerable attention revealing the possibility of investigating inaccessible gravitational phenomena in generic rotating geometries and flows by testing them through table top experiments. 
%This field of research owes its birth to Unruh who in 1981 showed that analogue Hawking radiation was present near wave horizons in a hydrodynamical fluid flow.
% Since then, many different analogues have been proposed including Hawking radiation, boson stars, and superradiance, and in a variety of physical systems ranging from nonlinear optics and Bose-Einstein condensates to hydrodynamics. 
Since the proposal to study analogue Hawking radiation in hydrodynamics \cite{Unruh1981}, many different astrophysical phenomena have been proposed: Hawking radiation, boson stars and superradiance analogues have been investigated in a variety of fields of physics, ranging from nonlinear optics to Bose-Einstein condensates (BECs) and hydrodynamics \cite{Liberati, FaccioBook,Garay_1,Garay_2,Schutzhold,Giovanazzi2004,Parentani2011,Steinhauer2014_1,Steinhauer2014_2,Faccio2010,Cardoso}. In this framework, the first measurement of analogue superradiance was reported in a recent study in classical fluid-dynamics where scientists were able to measure over-reflection of waves carrying Orbital Angular Momentum (OAM) scattered from a rotating vortex in a water tank \cite{Torres2017}. 
%All the analogues of superradiance provided until now provide a generalized framework for superradiant scattering in terms of Bogoliubov excitations.
In the last few years, several proposals have extended the concept of superradiance from classical fluids to superfluids, providing a generalized framework for superradiant scattering in terms of Bogoliubov excitations \cite{Marino2008_1,Marino2009,Prain2019,Solnyshkov2019}. A range of studies have focused on superfluids or photon fluids realised with light, i.e. with an optical beam propagating in a medium with a defocusing nonlinearity that mediates the background repulsive photon-photon interaction. This can be tailored so as to reproduce superfluid dynamics and physical phenomena ranging from shock dynamics to analogue black holes \cite{Rica,Chiao1,Wan07,Conti07,Carusotto2012,Elazar2012,Carusotto2014,CarusottoLarre2015,Vocke2015,Vocke16,GlorieuxPRL2018,Vocke2018}.\\
%In these systems, quantum pressure (healing length in BECs or diffraction in optics) is treated as an obstacle as it introduces a mathematical complication that undermines the otherwise direct  geometrical space-time analogy \cite{Prain2019,Solnyshkov2019}. \red{is this true? And Angus did NOT include quantum pressure? so how do we include it in our model and claim that we have a Kerr BH analogue?}\\
%
In this Letter we propose and analyze nonlinear optics as a new and flexible platform for the study and elaboration of Penrose superradiance.  Specifically, we consider a geometry in which a weak signal or probe beam carrying OAM is incident along with a strong vortex pump beam onto a nonlinear defocusing medium. Four-wave mixing (FWM) in turn generates idler modes that can be trapped in the core of the pump vortex via nonlinear cross-phase modulation.  Under suitable conditions the trapped modes can have negative norm, in which case the reflected signal power can be amplified, this being Penrose superradiance.  We first derive the conditions for amplification in terms of phase-matching familiar from nonlinear optics, and then relate this to the physics of Penrose superradiance.  Our analysis highlights how the concepts of ergoregion and positive and negative norm modes appear naturally in the nonlinear optics context. This extends the study of Penrose superradiance to the context of superfluidity, whilst providing also direct access to both the inner and outer regions of the vortex, thus providing insight for example into the transient dynamics of the process.\\
{\emph{Basic model and equations.}}
We consider the interaction between a continuous wave monochromatic pump field $E_0$ with OAM $\ell$  and a weak probe signal $E_s$ with OAM $n$ as described by the Nonlinear Schr\"odinger Equation (NSE). This is solved for the total light field $E=E_0+E_s+E_i$, where $E_i$ is the `idler' field generated by degenerate FWM, with idler OAM $q=2l-n$.  The NSE for the system is given by
\begin{equation} \label{fullNLS}
i\frac{\partial E}{\partial z} + \frac{1}{2k} \nabla_{\bot}^2 E - \frac{k n_2}{n_0}|E|^2 E = 0,
\end{equation}  
\noindent where $n_0$ is the linear refractive index, $k=2\pi n_0/\lambda$ is the wave-number, $\nabla_{\bot}^2$ is the transverse Laplacian accounting for optical diffraction, and we consider a defocusing medium with nonlinear coefficient $n_2<0$.

For co-propagating fields along the z-axis, the total field $E$ may be written in cylindrical coordinates $(r,\theta,z)$ as
\begin{eqnarray}
E(r,\theta,z) &=& E_0+E_s+E_i \label{line2} \\
&=& \left [ {\cal E}_0(r)e^{i\ell\theta} + {\cal E}_s(r,z) e^{in\theta} + {\cal E}_i(r,z)e^{iq\theta} \right ]e^{i\beta_\ell z},\nonumber %\label{line2}
\end{eqnarray}
where $\beta_\ell=k_0n_2I_\ell < 0$, with $I_\ell$ the background intensity of the strong pump vortex.  Then in the linearized regime the signal and idler fields are governed by the coupled equations
\begin{eqnarray}
{\partial {\cal E}_s\over\partial z} &=& {i\over 2 k} \nabla_n^2 {\cal E}_s + ik_0n_2\left [  2|{\cal E}_0|^2 {\cal E}_s+{\cal E}_0^2 {\cal E}_i^* \right ] -i\beta_\ell {\cal E}_s,  \label{CMT1}\\
{\partial {\cal E}_i\over\partial z} &=& {i\over 2 k} \nabla_q^2 {\cal E}_i + ik_0n_2\left [  2|{\cal E}_0|^2 {\cal E}_i + {\cal E}_0^2 {\cal E}_s^*\right ] -i\beta_\ell {\cal E}_i , \label{CMT2}
\end{eqnarray}
obtained by linearising Eq.~(\ref{fullNLS}) in the signal and idler fields and separating them on the basis of their different OAM (see Supplementary Information (SI) \cite{supply}).\\
It is important to note that viewed as a photon fluid our system is two-dimensional in the plane transverse to the direction of propagation. This is reflected in the fact that the effective frequency shifts for the signal (s) and idler (i) fields $ \Delta\omega^{s,i}=(\omega^{s,i}-\omega_p)=-c\Delta K_z^{s,i}/n_0$ that follow from Eqs. (\ref{CMT2}) correspond to phonon frequencies i.e. oscillation frequencies in the transverse plane {\cite{Vocke2015}}. \\
\emph{Ergoregion} -- We consider two characteristic speeds: the underlying flow speed $v$ and the speed of sound $c_s$ of the photon fluid, i.e. transverse perturbation modes. The ergoregion is defined as the region in the $(x,y)$ plane where $v>c_s$. In photon fluids, the speed of sound is defined as $c_s=(c/n_0)\sqrt{|\Delta n|/n_0 }$, where $\Delta n=n_2I_{\ell}$ is the nonlinear change in refractive-index due to the pump intensity $I_{\ell}$ \cite{Vocke2015}. The flow speed is $v= |\Omega|r = (c/n_0) \frac{|n-\ell|}{kr}$ where $\Omega$ is the pump rotational frequency with respect to the perturbation. Equating $c_s=v$ at the ergoradius yields $r_e=\frac{|n-\ell|}{k}\frac{n_0}{|\Delta n|}$. \\%We will see in the following that the interaction between the signal and the idler is maximum at $r_e$.\\
\emph{Positive and negative modes and currents} -- Penrose superradiance is based on the concept of positive and negative energy modes: negative energy modes  can remain trapped within the ergoregion, allowing positive energy modes to escape, gaining energy. It has been shown \cite{Carusotto2014,Prain2019} that Eqs.~(\ref{CMT1}-\ref{CMT2}) exhibit a conserved quantity $N(z)$, also referred to as a Noether current which, in our system, corresponds to
$J^0(r,z)$, where  $\partial_z  J^0 = 0$ and $J_0=|E_s|^2 - |E_i|^2$, such that
\begin{equation} \label{eq: RandTcoeffs}
N(z)=\int_{0}^{\infty} \bigl( |E_S|^2 - |E_I|^2 \bigr) rdr= \mbox{const}.
\end{equation}
%
%This quantity in BECs is usually proportional to the energy of the eigestate of the CMT equations by the relation $\mathbb{E}=\hbar\omega N(z)$, where $\omega$ is the mode frequency. It is worth noticing that this quantity is not positive-definite hence the energy of the eigenstate can be negative. 
Negative norm modes in our system arise from the idler wave intensity $|E_i|^2$. From Eq.~(\ref{eq: RandTcoeffs}) it is possible to also define the reflection, $R$, and transmission, $T$, coefficients for the modes scattering from the ergoregion:  
$R(z) = \int_{r_e}^{\infty} \left( |E_S|^2 - |E_I|^2 \right) rdr$ and 
$T(z) = \int_{0}^{r_e}      \left( |E_S|^2 - |E_I|^2 \right) rdr$. \\
%For normalized fields $R+T=1$, hence an increase in the amplitude of the reflected mode causes a decrease in the transmission coefficient. 
Superradiance results in a reflection coefficient larger than $1$, such that the reflected field has gained energy (or has been over-reflected) after scattering with the rotating body. %These coefficients being strongly related to the current distribution in the $(x,y)$ plane, 
The current $J_0$ is therefore a key signature for establishing the presence of superradiance, which can be identified by the presence of negative current $(J_0<0)$ inside the scattering region $r_e$, balanced with a positive current $(J_0>0)$ outside $r_e$ \cite{Prain2019}. %This reflects Penrose's physics based on the presence of negative energy modes inside $r_e$. 
Moreover,  a negative current inside $r_e$ implies that the idler wave has become trapped inside the ergoregion, while the signal is scattered outwards. %If there is no trapping of the idler, then the positive and negative modes can interfere and change the resulting energy outcome. 
In real black holes the trapping can be provided by the event horizon or by the ergosphere itself. \\%We will see that in our system, trapping is achieved via cross phase-modulation from the strong pump to the idler through the medium nonlinearity, that can create an effective waveguide in the core region.\\
\emph{Trapping of the idler wave} -- %Here we derive a condition for the Penrose process in a photon fluid. To proceed we first look at the propagation of a strong signal wave according to Eq.~(\ref{CMT1}), evaluate the trapped idler waves, and finally derive a condition (phase-matching) under which the trapped idler waves may be excited by the incoming signal wave, which gets amplified.
% \emph{Signal propagation} -- 
%For this analysis we assume that the signal is strong and negligibly depleted by the nonlinear interaction. In this limit 
Neglecting any effect of the idler on the signal propagation to lowest order in Eq.~(\ref{CMT1}), and assuming that the signal beam is not too tightly focused, we take the signal field as a focused Laguerre-Gaussian (LG) beam with radial mode index $p=0$, OAM $n$, and focused spot size $w_0$:
\begin{equation}\label{E_sig}
{\cal E}_s(r,z) \approx c_s V_n(r,z) e^{-i(1+|n|)\phi_G(z)}e^{2i\beta_\ell\Gamma_n(z) z-i\beta_\ell z},
\end{equation}
where $V_n(r,z)$ is the normalized z-dependent LG mode profile,  we have explicitly separated out the Gouy phase-shift that occurs through the beam focus with $\phi_G(z)=\tan^{-1}(z/z_0)$, and the Rayleigh range is $z_0=kw_0^2/2$. Here $\Gamma_n(z) = \int_0^\infty 2\pi rdr~|V_n(r,z)|^2 u_\ell^2(r) $ describes the variation of the signal phase due to penetration of the LG mode into the pump vortex core. From Eq.~(\ref{E_sig}) we can write the signal wavevector (nonlinear) shift as
\begin{equation}
\Delta K_s \approx \Delta K_s(0) = 2\beta_\ell\Gamma_n(0) -\beta_\ell,
\end{equation}
where we accounted for the fact that most of the nonlinear interaction will occur within a Rayleigh range around the beam focus at $z=0$. The overlap factor $0\le\Gamma_n(0)\le 1$ may be evaluated numerically. At the focus, the radius of the single-ringed LG beam is $r_n = w_0 \sqrt{n/2}$. Moreover, we require $r_n \approx r_e$ in order for the signal LG ring beam to glance off the ergosphere at it goes through its focus.\\
We now consider the idler propagation according to Eq.~(\ref{CMT2}). This can be re-arranged as
\begin{equation}\label{IdlEq}
{\partial {\cal E}_i\over\partial z} ={i\over 2k} \nabla_q^2 {\cal E}_i + i\underbrace{2\beta_\ell [u_\ell^2(r) -1]}_{waveguide} {\cal E}_i  + i\beta_\ell {\cal E}_i +\underbrace{ i\beta_\ell u_\ell^2(r) {\cal E}_s^* }_{source}. 
\end{equation}
such that it is composed of two terms: (i) a two-dimensional waveguide term, $2|\beta_\ell| [1-u_\ell^2(r)]=k_0\Delta n(r)$, that arises from the cross-phase-modulation induced by the pump vortex on the idler wave and, (ii) a source term describing how the idler wave (absent at the input) is driven by the signal beam via the FWM interaction. 
To proceed it is useful to assess the idler guided modes $U_{pq}(r)$ that arise in the presence of the waveguide term in Eq. (\ref{IdlEq})  while ignoring the source term
\begin{equation}\label{E_idl}
{\cal E}_i(r,z) = c_i(z) U_{pq}(r)e^{i(\beta_\ell +\Lambda_{pq})z} ,
\end{equation}
with radial mode-index $p$, eigenvalue $\Lambda_{pq}$, and idler wavevetor $\Delta K_i = \beta_\ell + \Lambda_{pq}$. In order to verify the existence of guided modes, we compute the spectrum of the idler waves with OAM $q$ for a given pump vortex profile $u_\ell(r)$ and value of the nonlinear parameter $\beta_\ell$ (see SI \cite{supply}).  We can then find a condition for the incident signal field to excite a guided idler mode by substituting the guided idler field of the form of Eq.~(\ref{E_idl}) into Eq.~(\ref{CMT2}), with the signal field in Eq.~(\ref{E_sig}), giving
\begin{equation}\label{dcidz}
{dc_i\over dz} = ic_s^* \beta_\ell F(z) \underbrace{ e^{-i(2\Delta K z - (1+|n|)\phi_G(z))} }, 
\end{equation}
where
\begin{equation}
\begin{split}
\Delta K & = \left ( {\Delta K_s + \Delta K_i \over 2} \right ),\\
F(z) &=  \int_0^\infty 2\pi rdr~V_n^*(r,z)u_\ell^2(r) U_q^*(r). 
\end{split}
\end{equation}
It is possible to solve Eq.~(\ref{dcidz}) numerically to explore how effectively the signal excites the idler guided mode for a given set of parameters, but the main insight can be gained by looking at phase-matching conditions dictated by the underbraced exponential phase factor. In the vicinity of the origin, the phase factor is approximately $(2\Delta K z - (1+|n|)z/z_0)$, so that 
 \begin{equation}
\Delta K = (2\beta_\ell\Gamma_n(0) -\beta_\ell )
 + ( \beta_\ell + \Lambda_{q}).
 \end{equation}
If the Gouy phase-shift term is zero ($\phi_G=0$), then $\Delta K=0$ for phase-matching and efficient generation but a more general condition  $\Delta K>0$ guarantees the possibility of phase-matching. Indeed, similar phase factors as in Eq.~(\ref{dcidz}) along with the $\Delta K>0$ condition appear in the theory of harmonic generation using focused beams \cite{BoydBook}.
In our system, the $\Delta K>0$ condition can be used to determine whether the guided idler waves can be excited, with consequent observation of Penrose superradiance. \\%Indeed, if $\Delta K<0$ the Penrose process cannot arise.
\emph{Zel'dovich-Misner condition} -- The condition $\Delta K > 0$ can be recast in terms of transverse perturbation (i.e. phonon) frequencies as  $\Delta \omega=-(c/n_0)\Delta K < 0$, where $\Delta\omega = \left ({\Delta\omega_s+\Delta\omega_i\over 2} \right )$ is the average of the frequency shifts of the signal and idler fields, with $\Delta\omega_{s,i}=-(c/n_0)\Delta K_{s,i}$.  We note that $\Delta\omega=(\omega - \omega_p)=(\omega - m\Omega)$, so the condition to see Penrose superradiance is $(\omega - m\Omega) < 0$. This has the same form as the Zel'dovich-Misner condition \cite{Zeldy1971,Misner1972}, therefore establishing the connection between the nonlinear dynamics of the optical beams and the cornerstone relation for Penrose superradiance.
 
\emph{Numerical simulations} --
To quantitatively study our proposed nonlinear optics platform for Penrose superradiance, we numerically simulate Eqs.~(\ref{CMT1}-\ref{CMT2}) in a defocusing nonlinear medium.  We assume a strong vortex pump beam that does not vary with propagation distance $z$, and neglect absorption.  To reveal the generic nature of our results we employ dimensionless variables, where $x,y$ and $r$ are expressed in units of the signal spot size $w$, and $z$ is in units of the Rayleigh range $Z_R = kw^2/2$.  The background fluid vortex is generated by a vortex pump background, $E_0(r) = N_0 \tanh(r)^{|l|}e^{il\theta}$,
where $\ell$ is the vortex charge and $N_0$ is a normalization constant. This form for the pump is an excellent approximation to the exact solution of the NSE in Eq. (\ref{fullNLS}) {\cite{Dinev1997}}.  The input signal field is a Laguerre-Gauss beam with OAM $n$, and is written 
\begin{equation}
E_s(r,z=0) = N_s \left({r\over w}\right)^{|n|}e^{-{r^2\over w^2}}e^{in\theta},
\end{equation}
where $N_s$ is a normalization constant, and the idler beam is chosen to have zero amplitude at the input.\\
\begin{figure*}[t] 
\centering
\includegraphics[width=2\columnwidth]{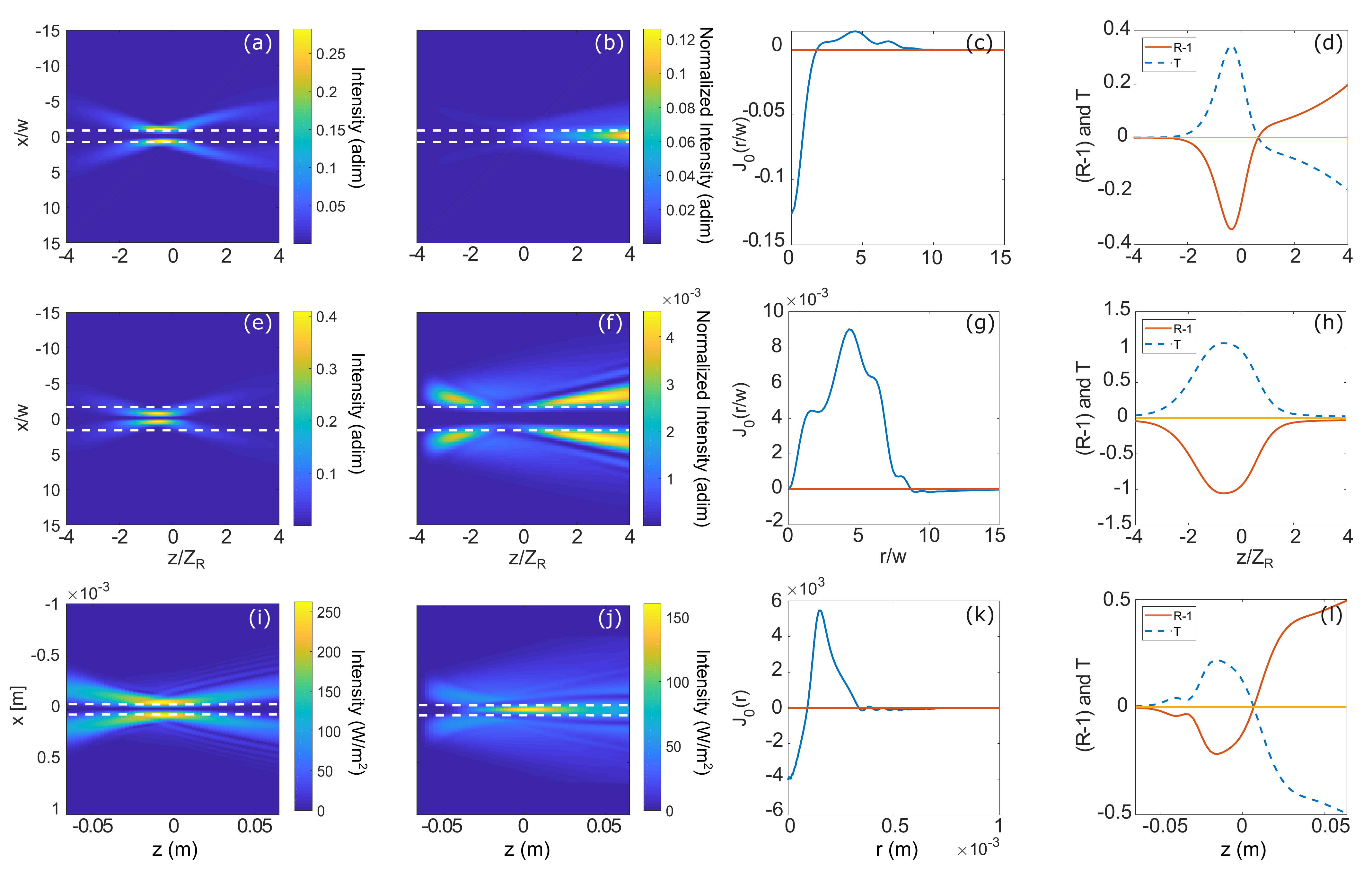}
\caption{\small{For all cases the pump has OAM $\ell=1$. (a-d) For this case the signal has OAM $n=2$ and idler OAM $q=0$, and Penrose superradiance occurs, $\Delta K>0$.  Cross-sections for the intensity profiles $|E_{s,i}(x,y,z)|^2$ versus $x/w$ and $z/Z_R$ for $y=0$ for the signal (a) and idler (b), calculated using the linearized theory:  (c) Current $J_0(r/w)$ versus radius $r/w$ for $z/Z_R=4$, and (d) reflection $(R-1)$ and transmission $T$ versus propagation distance $z/Z_R$.  (e-h) As in panels (a-d) for signal with OAM $n=-1$ and idler with OAM $q=3$ for which Penrose superradiance is absent, $\Delta K < 0$. (i-l) as in (a-d) calculated using the full NSE (\ref{fullNLS}). The horizontal dashed white lines in panels (a-b,e-f) indicate the location of the dimensionless radius $r/w$ of the ergoregion}}
\label{fig:ewan_sim}
\end{figure*}
Figures~\ref{fig:ewan_sim}(a-h) show the signal and idler evolution for two cases with pump OAM $\ell=1$: (1) Figs.~\ref{fig:ewan_sim}(a-d) - in this case $\Delta K>0$ and the signal %with OAM $n=2$ 
undergoes superradiant amplification while the idler %with OAM $q=0$ 
remains trapped inside the ergoregion; (2) Figs.~\ref{fig:ewan_sim}(e-h) - in this case $\Delta K<0$ and superradiance does not occur.   The horizontal dashed white lines in Figs.~\ref{fig:ewan_sim}(a-b,e-f) indicate the location of the dimensionless radius $r/w$ of the ergoregion.\\
In more detail, for Figs.~\ref{fig:ewan_sim}(a-d) the signal with OAM $n=2$ intensity profile $|E_s(x,y=0,z)|^2$ versus $z/Z_R$ is shown in (a) along with the corresponding idler intensity with OAM $q=0$ in (b). We see that the idler field is initially absent and generated via FWM during propagation. As the signal diffracts away from the ergoregion for $z/Z_R > 0$ it experiences a transient amplification, in contrast to the idler beam that remains trapped inside the waveguide created by the pump vortex core. We also note that the idler is first generated outside the ergoregion for $z/Z_R< 0 $, and only past the signal focus the idler becomes trapped.  In the context of Penrose superradiance, we interpret this as evidence that negative energy waves are trapped during the interaction, while positive energy waves are reflected (no signal inside the ergoregion). This is further confirmed by the plot of the current $J_0(r/w)$ versus $r/w$ at $z=4$ in (c). We see that within the ergoregion the current is negative as implied by our theoretical analysis. Plot (d) shows the reflection (R-1) and transmission (T) coefficients, defined in Eq.~(\ref{eq: RandTcoeffs}), indicating an amplification of $19\%$ at $z=4$.\\
Figures~\ref{fig:ewan_sim}(e-h) show the corresponding results for the case with no superradiance, $\Delta K<0$. Here the signal OAM is $n=-1$ giving OAM $q=3$ for the idler. We note that the current $J_0(r/w)$ is now positive near the origin at $z/Z_R=4$ as shown in plot (g), and that the reflection coefficient is $(R-1)\simeq -2\%$ as follows from plot (h), meaning that there is no amplification in this case.  Furthermore, the idler is no longer trapped inside the ergoregion as illustrated in plot (f), the signal intensity profile being shown in plot (e).  These results illustrate the fact that trapping of a negative mode inside the ergoregion and amplification of the signal go hand-in-hand, as expected  for Penrose superradiance. \\
%
%\emph{NLS Simulations} -- 
In order to prove the presence of superradiance in a real photon fluid experiment, we finally simulate the interaction with the full NSE (\ref{fullNLS}) in a defocusing nonlinear medium. The sample parameters are chosen based on previous experiments in photon fluids experiments with linear refractive index is $n_0=1.32$,  and nonlinear refractive index $n_2=1.2 \times 10^{-10}$ m$^2/$W. The nonlinearity is assumed to be local as in experiments with Rb atoms \cite{GlorieuxPRL2018} or in methanol and graphene solutions where time-gated measurements allow to choose a nonlocal length shorter than healing length \cite{Vocke2018}.\\
%\begin{figure*}[t] 
%\centering
%\includegraphics[width=0.8\textwidth]{Figs/exp_sim_NLS_SR2.eps}
%\caption{\small{The pump field $E_0$ has OAM $\ell=1$. (a,b) Section at $y=0$ of the signal (a) and idler (b) fields intensities evolution. Signal OAM $n=2$ and Idler OAM $q=0$. (c) Current $J_N(r)$ as function of the radius $r$ computed after Eq. (\ref{eq: RandTcoeffs}) at $z=13$cm. Red line shows the null current axis. The dashed line marks the location of the ergoregion $r_E$. (d-g) as in panel (a-c) for signal with OAM $n=3$ and idler with OAM $q=-1$. Panel (g) is the zoom of the red square area marked in panel (f). The dot-dashed red line shows the location of the waveguide induced radius. (h-k) as in (d-g) with $n=-1$ and $q=3$. (l-o) as in (d-g) with $n=4$ and $q=-2$.}}
%\label{fe5}
%\end{figure*}
%
The initial field is a beam at wavelength $\lambda=532$ nm given by the superposition of a pump super-Gaussian vortex $e^{-(r/w_0)^{10}}\mbox{tanh}^{|\ell|}(r/w_v)e^{i\ell\theta}$ with $\ell=1$, and a Laguerre-Gaussian probe signal $(r/w_v)^{|n|}e^{-(r/w_v )^2}e^{in\theta}$ with $n=2$.
%The radial phase on the probe beam is $e^{-ik_rr}$ where $k_r=k_0sin(\alpha)$ and $\alpha=0.06^{\circ}$ is the tilt angle, $k_0=2 \pi n_0/\lambda$. 
The pump power is chosen as $140$~mW as in \cite{Vocke2018}, several orders of magnitude higher than the signal beam $P_{s}=10^{-2}P_{pump}$.  Figures~\ref{fig:ewan_sim}(i-l) in the bottom row show the same quantities as the top row for the chosen parameters, and the results shows all of the main features of Penrose superradiance, including a negative current near the origin in plot (k), trapping of the idler beam in plot (j), and amplification up to $50\%$ in plot (l).  The relevance of these results is that they no longer rely on the linearized approximation or the assumption that the pump vortex does not evolve with propagation distance, and so illustrate that the signatures of Penrose superradiance can survive the unavoidable imperfections of a real experiment.  Furthermore, the parameters employed and the scales involved show that an our nonlinear optics platform for studying Penrose superradiance is entirely feasible.\\%
\emph{Conclusions} --
%
%Despite the long history of Penrose superradiance and predictions that this should occur in a variety of settings beyond black holes, this has so far only been observed in a water vortex experiment. % and has not yet been observed in nonlinear optics and in superfliuds due to the lack of a transient description of the phenomenon and the stability of the system. 
We have provided theory and simulations to prove that Penrose superradiance can also be explored in the regime of superfluid light, and we have connected the main physical concepts of the Penrose process to the related counterparts in nonlinear optics.  Our analysis generalizes the definition of superradiance by showing that this can be readily observed by characterising the currents across the ergoregion, and highlights the key role played here by diffraction and mode-trapping. The results  imply a new amplification regime in nonlinear optics that is tightly connected to the trapping of the idler beam that, when spatially separated from the signal beam, creates an effective  loss, leading in turn to a transient gain for the signal beam. This bears a close connection to non-normal dynamics in a coupled resonator system in the presence of loss in one of the resonator modes \cite{Pol}.  These results pave the way towards future experiments on superradiant amplification in nonlinear optics and a deeper understanding of the fundamental physics and transient dynamics of Penrose superradiance. \\% Due to the experimental feasibility of our proposal, we expect an experimental proof of superradiance in superfluids to come soon.

\emph{Acknowledgements} --
The authors acknowledge financial support from EPSRC (UK Grant No.
EP/P006078/2) and the European Union's Horizon 2020
research and innovation programme, grant agreement
No. 820392.

%\bibliography{references}
%

\newpage

\onecolumngrid
{\centering{\large \bfseries Penrose Superradiance in a Nonlinear Optics Superfluid: supplementary information\par}\vspace{2ex}
	{Maria Chiara Braidotti$^{1}$, Daniele Faccio$^{1,2}$, Ewan M. Wright$^{2}$\par}}
{\centering  \small \emph{$^{1}$School of Physics and Astronomy, University of Glasgow, G12 8QQ, Glasgow, UK.\\ 
$^{2}$Wyant College of Optical Sciences, University of Arizona, Tucson, Arizona 85721, USA.\\ }\par}
\smallbreak

\par\vspace{1ex}

\renewcommand{\theequation}{S\arabic{equation}}
\renewcommand{\thefigure}{S\arabic{figure}}
\setcounter{equation}{0}

\section{Linearized equations and theory}

In this section we derive the linearized Eqs. (3-4) of the main text that govern the evolution of the signal and idler fields in the photon fluid in the presence of the strong pump field. We start from the nonlinear Schr\"odinger equation for a monochromatic field of frequency $\omega_0$ and a local nonlinearity describing propagation in the photon fluid
\begin{equation}\label{Eq1}
{\partial E\over\partial z} = {i\over 2k}\nabla_\perp^ 2 E + ik_0n_2 |E|^2 E ,
\end{equation}
where $E$ is a monochromatic light field with wavelength $\lambda$, $n_0$ the linear refractive index of the medium, $k=2\pi n_0/\lambda=n_0k_0=n_0\omega_0/c$ is the wavenumber, and $\nabla_{\bot}^2$ is the transverse Laplacian, accounting for optical diffraction. We observe that optical fluids are two-dimensional , that is, they live in the plane $(x,y)$ orthogonal to the direction of propagation $z$. We consider a defocusing nonlinearity $(n_2<0)$ and a vortex pump solution of the form
\begin{eqnarray}\label{E0}
E(r,\theta,z) &=& {\cal E}_0(r) e^{i(\beta_\ell z + \ell\theta)} \nonumber \\
&=& \sqrt{I_\ell} ~u_\ell(r)  e^{i(\beta_\ell z + \ell\theta)} ,
\end{eqnarray}
where $I_\ell$ is the background intensity of the vortex of OAM $\ell$, $u_\ell(r)$ is the corresponding vortex profile which has a core size denoted $r_\ell$, and $\beta_\ell=k_0n_2I_\ell < 0$.  The vortex profile, which we take as real without loss of generality, obeys the equation
\begin{equation}
\beta_\ell u_\ell = {1\over 2k}\nabla_\ell^ 2 u_\ell + k_0n_2 I_\ell u_\ell^3  ,
\end{equation}
where $u_\ell(r)\rightarrow 1$ for $r>>r_\ell$, and we have defined $\nabla_p^2 = {\partial^2\over\partial r^2} + {1\over r}{\partial\over\partial r} -{p^2\over r^2}$.

To proceed we make use of the fact that in the presence of the strong pump ${\cal E}_0$ of OAM $\ell$ and a weak externally applied signal field ${\cal E}_s$ of OAM $n$, the total field may be written as
\begin{eqnarray}
E(r,\theta,z) &=& \left [ {\cal E}_0(r)e^{i\ell\theta} + {\cal E}_s(r,z) e^{in\theta} + {\cal E}_i(r,z)e^{iq\theta} \right ]e^{i\beta_\ell z}, \label{line1} \\
&=& \left [ {\cal E}_0(r) + {\cal E}_s(r,z) e^{i(n-\ell)\theta} + {\cal E}_i(r,z)e^{-i(n-\ell)\theta} \right ]e^{i(\beta_\ell z+\ell\theta)},
\end{eqnarray}
with ${\cal E}_i$ the generated idler field with OAM $q=(2\ell-n)$.  %Note that according to the first line longitudinal wavevector shifts $\Delta K_z$ calculated from the field envelopes ${\cal E}_{s,i}$ are automatically referenced to the pump value $\beta_\ell$, so that all corresponding frequency shifts $
%\Delta\omega=(\omega-\omega_p)=-c\Delta K_z/n_0$ are with respect to the associated pump frequency $\omega_{p}=-c\beta_\ell/n_0$ (the negative sign is a consequence of the different ways $z$ and $t$ enter the plane-wave factor $e^{i(\Delta K_z z-\Delta\omega t)}$ for a traveling wave). {\it Here we stress that the wavevectors $\Delta K_z$ and the associated frequencies $\Delta\omega$ shifts are phononic in nature, that is, they apply to the phonons of the photon fluid}.  Likewise, in the second line the OAM of the signal $(n-\ell=m)$ and idler $(q-\ell=\ell-n=-m)$ fields are referenced to the pump value $\ell$.  Referencing the phononic wavevectors and frequencies, along with the OAM of the signal and idler fields, to the pump values is key in applying the ideas of Ref. \cite{Prain2019} to our photon fluid system.

Then substituting the expansion (\ref{line1}) into the starting Eq. (\ref{Eq1}), linearizing in the signal and idler fields, and separating the signal and idler equations on the basis of their differing OAM yields
\begin{eqnarray}
{\partial {\cal E}_s\over\partial z} &=& {i\over 2 k} \nabla_n^2 {\cal E}_s + ik_0n_2\left [  2|{\cal E}_0|^2 {\cal E}_s+{\cal E}_0^2 {\cal E}_i^* \right ] -i\beta_\ell {\cal E}_s, \nonumber \\
{\partial {\cal E}_i\over\partial z} &=& {i\over 2 k} \nabla_q^2 {\cal E}_i + ik_0n_2\left [  2|{\cal E}_0|^2 {\cal E}_i + {\cal E}_0^2 {\cal E}_s^*\right ] -i\beta_\ell {\cal E}_i . 
\end{eqnarray} 
Finally we substitute the strong vortex pump in Eq. (\ref{E0}) into the above equations to obtain
\begin{eqnarray}\label{sigidl}
{\partial {\cal E}_s\over\partial z} &=& {i\over 2 k} \nabla_n^2 {\cal E}_s + i\beta_\ell u_\ell^2(r) \left [  2{\cal E}_s+{\cal E}_i^* \right ]  -i\beta_\ell {\cal E}_s,\nonumber \\
{\partial {\cal E}_i\over\partial z} &=& {i\over 2k} \nabla_q^2 {\cal E}_i + i\beta_\ell u_\ell^2(r) \left [  2{\cal E}_i + {\cal E}_s^*\right ]  -i\beta_\ell {\cal E}_i. 
\end{eqnarray} 
These equations are the basis for the manuscript discussion and describe the parametric interaction between the signal and idler fields in the presence of the pump, this parametric interaction arising from Four Wave Mixing (FWM). 

%Our simulations are performed using an incident focused Laguerre-Gaussian (LG) signal beam: The goal is that the signal LG ring beam is configured to glance off the core of the pump vortex at focus  Four-wave-mixing then produces an idler field, and the Penrose process can cause some of the idler power to become trapped within the central ergosphere while the signal dominantly expands away from this region.  The idea is that this is analogous to the situation in which for an e-p pair created near a black hole ergosphere one of the pair can fall into the black hole while the other escapes.

\section{Trapping of the idler wave}

\subsection{Idler propagation}
The idler propagation equation (\ref{sigidl}) can be rearranged as
\begin{equation}\label{IdlEq}
{\partial {\cal E}_i\over\partial z} ={i\over 2k} \nabla_q^2 {\cal E}_i + i\underbrace{2\beta_\ell [u_\ell^2(r) -1]}_{waveguide} {\cal E}_i  + i\beta_\ell {\cal E}_i +\underbrace{ i\beta_\ell u_\ell^2(r) {\cal E}_s^* }_{source}. 
\end{equation}
Since the nonlinear parameter $\beta_\ell$ is negative, the underbraced term $2|\beta_\ell| [1-u_\ell^2(r)]=k_0\Delta n(r)$ defines a two-dimensional refractive-index profile which is guiding since $u_\ell^2(r)$ is zero at the pump vortex center $r=0$, so $\Delta n(r)=2|\beta_\ell|$ is maximum there, and is unity for $r>>r_\ell$ away from the vortex core, so that $\Delta n(r)$ goes to zero. The pump vortex therefore creates a cross-phase-modulation (XPM) induced waveguide that is experienced by the idler wave. %(this also underpins the XPM-induced nonlinear phase shift $2\beta_\ell\Gamma_n(z)$ experienced by the signal in Eq. (\ref{E_sig})).

The underbraced source term in the above equation describes how the idler wave, that is absent at the input, is driven by the signal beam via the parametric interaction.  As shown in the manuscript (see Eq.~(6) of the manuscript) the signal field can be approximated as 
\begin{equation}\label{E_sig}
{\cal E}_s(r,z) \approx c_s V_n(r,z) e^{-i(1+|n|)\phi_G(z)}e^{2i\beta_\ell\Gamma_n(z) z-i\beta_\ell z},
\end{equation}
where $V_n(r,z)$ is the normalized z-dependent Laguerre-Gauss mode profile, $\phi_G(z)=\tan^{-1}(z/z_0)$ is the Gouy phase-shift at the focus with Rayleigh range defined as $z_0=kw_0^2/2$ and $\Gamma_n(z) = \int_0^\infty 2\pi rdr~|V_n(r,z)|^2 u_\ell^2(r) $ being the signal phase variation induced by the pump core on the signal.\\ 
In the following section we compute the guided idler modes spectrum. 
\subsection{Guided idler modes}
The spectrum of guided idler waves with OAM $q$ can be found by solving the wave equation (\ref{IdlEq}) combining beam diffraction and the XPM-induced refractive-index profile. Neglecting the source for the time being, and for idler fields of the form
\begin{equation}
{\cal E}_i(r,z) = c_i U_{pq}(r)e^{i(\beta_\ell +\Lambda_{pq})z} ,
\end{equation}
with radial mode-index $p$, this leads to the equation for the modes $(p=0,1,2\ldots)$
\begin{equation}
\left ( {1\over 2k} \nabla_q^2  + 2\beta_\ell [u_\ell^2(r) -1] \right ) U_{pq}(r) = \Lambda_{pq} U_{pq}(r).
\end{equation}
This eigenproblem can be solved for the guided idler modes for a given pump vortex profile $u_\ell(r)$ and value of the nonlinear parameter $\beta_\ell$. Note that it is possible that no guided idler modes exist in which case the Penrose process cannot occur.  The eigenvalues $\Lambda_{pq}$ are positive and decrease with increasing $p$, and for the present purposes the lowest radial mode $p=0$ is the relevant one. We hereafter drop the radial mode index for simplicity in notation, and assume $U_q(r)$ and $\Lambda_q$ exist and we obtain them numerically.

The modal solution $U_q(r)$ allows us to evaluate the ergosphere radius in a more systematic way:  this mode has a single-ringed intensity profile and one can find the radius $r_q$ of the peak intensity.  Physically any idler energy excited this guided mode will effectively be confined or trapped within the radius $r_q$, so we identify $r_q$ with a viable measure of the radius of the ergosphere $r_e$.  The approach $r_{e}=r_q$ agrees quite well with the previous approximation, particularly for larger $q$.

\end{document}